# Curvature Gradient Driving Droplets in Fast Motion

Cunjing Lv[1,2], Chao Chen[1,2], Yajun Yin[1], Fan-gang Tseng[3,4], Quanshui Zheng[1,2*]

[1] Department of Engineering Mechanics, Tsinghua University, Beijing 100084, China

[2] Center for Nano and Micro Mechanics, Tsinghua University, Beijing 100084, China

[3] Department of Engineering and System Science, National Tsing Hua University, Hsinchu 30013, Taiwan

[4] Research Center for Applied Sciences, Academia Sinica, Taipei 11529, Taiwan

**Earlier works found out spontaneous directional motion of liquid droplets on hydrophilic conical surfaces, however, not hydrophobic case. Here we show that droplets on any surface may take place spontaneous directional motion without considering contact angle property. The driving force is found to be proportional to the curvature gradient of the surface. Fast motion can be lead at surfaces with small curvature radii. The above discovery can help to create more effective transportation technology of droplets, and better understand some observed natural phenomena.**

Recently, spontaneous or self-motion phenomena of liquid droplets on solid surfaces have attracted a lot of attentions[1-5]. It was found that oil droplets of volume ~ 3 $\mu$L deposited on the surface of a conical fiber of diameters 0.1 mm spontaneously evolved into annual-shaped, and then moved towards the fiber's larger cross-section region[6]. In contrast, annual water or oil droplets tapered inside a conical tube were found to spontaneously move towards the tube's smaller cross-section region[7,8]. Theoretical analysis demonstrated that the surface tension can lead to the above phenomenon[7]. More recently, it was reported that spider silk is capable of collecting water efficiently from the air[5]. The observation shows a similar self-motion mechanism, namely, annual water droplets move on conical surfaces, although the authors noted another mechanism, known as the Marangoni effect[9], that the wettability may vary with conical axial direction.

---

[1]To whom correspondence should be addressed. E-mail: zhengqs@tsinghua.edu.cn

[2]The two authors Cunjing Lv and Chao Chen have equal contributions to this paper.

We notice that all the above researches are regarding annual droplets movement on hydrophilic conical surfaces. However, smaller droplets tend to be clam-shell-shaped[10] when deposited on conical surfaces with larger diameters. To the best of our knowledge, there is still lacking report on self-motion of clam-shell-shaped droplets on a conical surface, or surfaces with curvatures. Furthermore, we also wonder if a similar self-motion phenomenon of droplets would happen on a hydrophobic curved surface.

Motivated by the above the questions, we first carry out molecular dynamics (MD) simulations for water nanodrops on graphene cone based on the platform of LAMMPS[11]. In the simulations, the cone is treated to be rigid, and a widely used code, the SPC/E[12] model, is employed to describe the water with the same parameters given in Ref. 13. A Lennard-Jones potential, $u(x)=4\varepsilon[(\sigma/r)^{12}-(\sigma/r)^{6}]$, is used to characterize the carbon-water van der Waals interaction[14], where $r$ is the distance between a pair of carbon-oxygen atoms, $\varepsilon$ is the well depth of interaction, and $\sigma$ is related to the equilibrium distance of the carbon-oxygen pair. To model different wettabilities, we fix $\sigma = 0.319$ nm and choose $\varepsilon = 5.85$, 1.95, and 1.50 meV, with corresponding water contact angles of $\theta = 50.7°$ (hydrophilic)[15], 138° (hydrophobic), and nearly 180° (superhydrophobic), respectively. Here we emphasize that the "graphene" is used as a model, rather than a real material, in order to consider different wettabilities in a way as simple as possible. Incidentally, we note that the true contact angle of water on graphene surface is still unknown.

As illustrated in the inserts of Figure 1, we first choose a cone with the half-apex angle of $\alpha = 19.5°$ and cone height of $Y = 7$ nm from the cone tip, and then cut off the cone tip at height $Y = 1.5$nm or $Y = 3.5$nm, respectively, for modeling the movement of a water drop containing 339 atoms on the outer or inner conical surface. The diameter of this drop in spherical shape is estimated about 2nm. In the simulation, the temperature of water is kept at 300K with Nosé/Hoover thermostat[16,17], and the whole system is located in a finite vacuum box. At beginning, the initial mass center of the droplet is fixed to apply thermal equilibrium for 100 ps. Then the droplet is released to move freely along the conical surface for next 300 ps. The trajectory results along the meridian line are shown in Figure 1 and are discussed below.

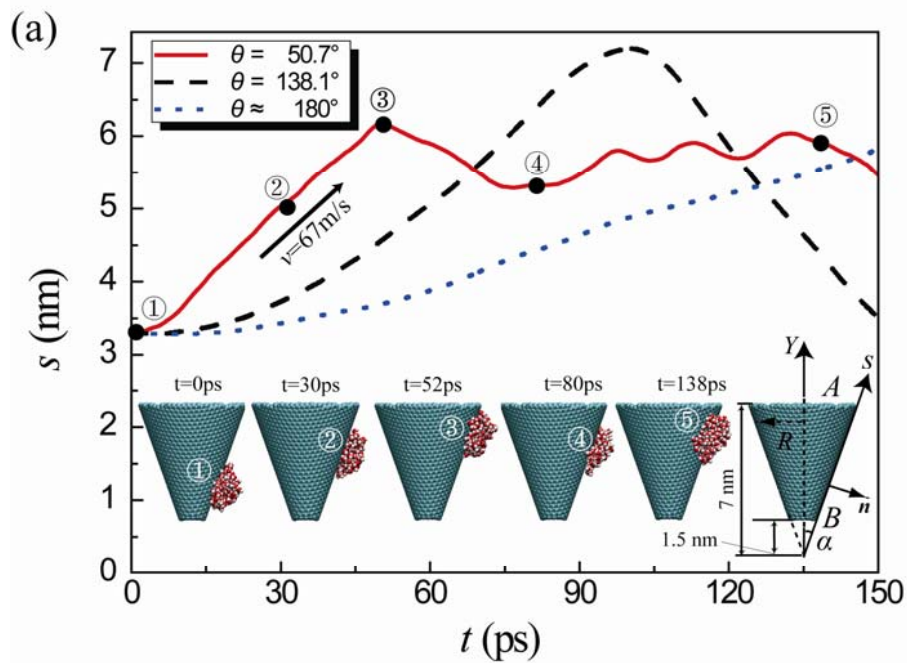

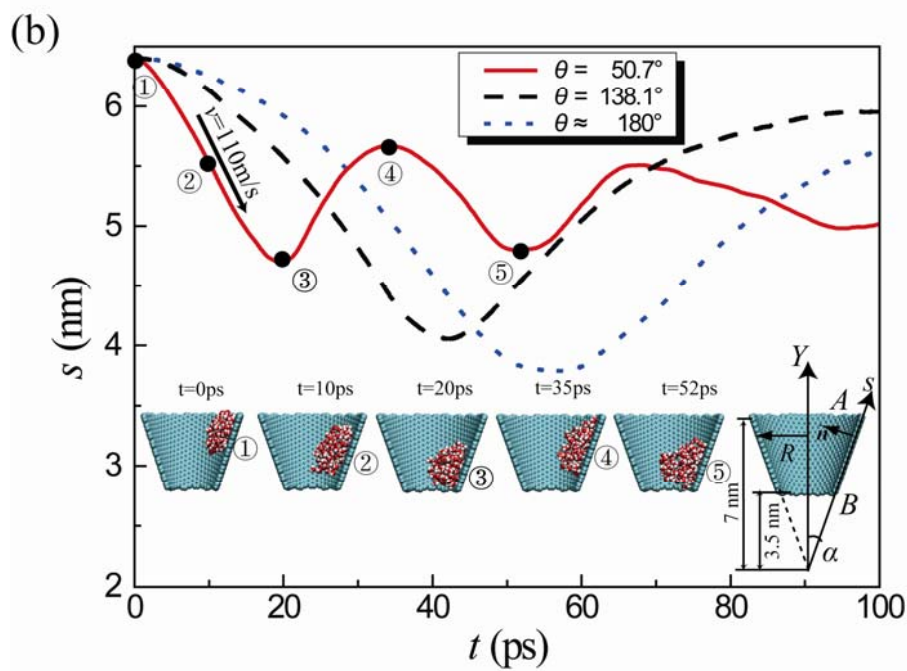

**Figure 1.** The self-moving behaviors of a nanodrop of 339 water molecules on the outer surface (a) and the inner surface (b) of the cone. The red solid, black dashed and blue dotted trajectory curves correspond to the water contact angles of 50.7°, 138° and nearly 180°, respectively. The inserts in (a) and (b) show the drop positions at the outer and inner surfaces of the cone respectively at five different moments selected from the MD simulation movie (see Supplementary Information).

When the droplet was released at the B position on the outer surface, the simulation results show that the droplet immediately started a self-motion toward the larger open end (A). After it met the end A, the droplet was bouncing back, and then moved toward the end B. $s$ is denoted as the distance between the projection point of the droplet mass center on the cone surface and the cone tip. The red solid curve in Figure 1(a) depicts the dependence of $s$ to the simulation time $t$ when the droplet moving at the outer surface with contact angle $\theta = 50.8°$, and the inserts show 5 selected frames from the movement movie (see Supplementary Information). Similar directional self-movement and bouncing were found on the hydrophobic (138°) and superhydrophobic (~180°) surfaces, as shown by the black dashed- and blue dotted-lines in Figure 1(a). When the water droplet was released at the inner surface, we observed diametrical results, as illustrated in Figure 1(b). The droplet started self-movement but toward the smaller open end (B). Similar bouncing phenomenon appeared when the droplet reaches the end B.

If we define the normal direction (**n**) of the surface to be toward the droplet (see the two rightmost inserts in Figure 1), then the two principal curvatures at distance $s$ are $k_1 = \pm \cos\alpha/R$ and $k_2 = 0$ as the droplet sets on the inner or outer conical surface, where $R$ is the radius of cone at distance $s$ that is equal to $s \cdot \sin\alpha$ (see the last insert of Figure 1(a)). Thus, from the above results we conclude that the self-motion direction is always toward the region with larger curvature, without considering the contact angle property. To our best of knowledge, this wettability-independence property is first reported. On the other hand, it can be also seen from Figure 1 that the smaller the contact angle of the surface has, the larger the self-motion speed of the droplet can reach. This rule is valid for the movement on both outer and inner conical surfaces.

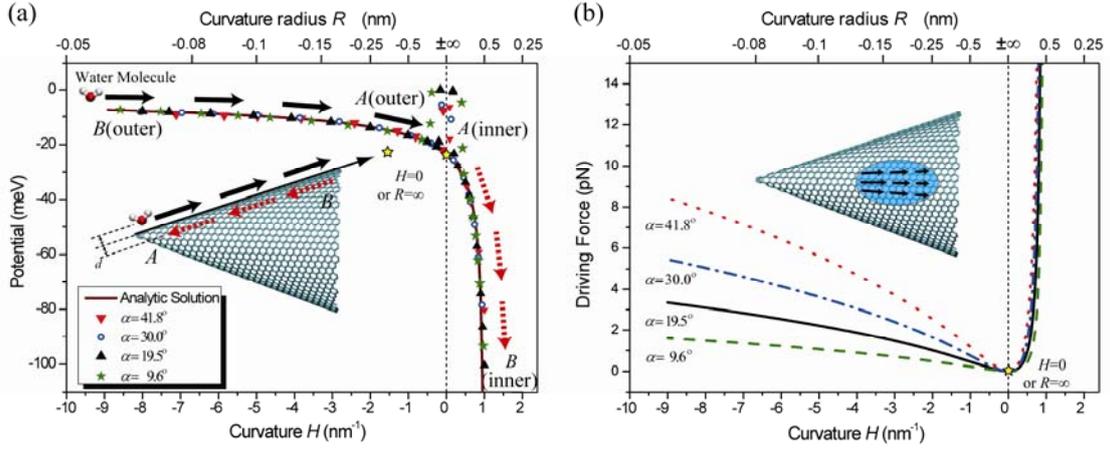

**Figure 2.** (**a**) The van der Waals potential $U$ between a single water molecule and all atoms of a conical graphene surface as the water molecule is kept at the constant separation $d = 0.5$nm to the surface. The four types of dots present four different half-apex angles, and the solid curve is the result of analytical solution Eq. (2). The inset is the schemtic of the movement path for the water molecule, while the black and red arrows present that the molecule on the outer or inner surfaces of the cone, respectively. (**b**) The driving force $-dU/ds$ as a function of the mean curvature ($H$) of the cone. Similar results are also obtained for other values of $d$.

To understand the above-reported self-motion phenomenon, we study the van der Waals potential energy, $U$, between a single water molecule and the conical graphene surface. We calculate $U$ by adding all the Lennard-Jones interaction energies between the oxygen atom of the water molecule and all carbon atoms of the graphene conical surface. The results of $U$ corresponding to the contact angle $\theta = 50.7°$, and the half-apex angle $\alpha = 19.5^0$, and a fixed distance $d = 0.5$nm between the oxygen and the conical surface are plotted as black upper triangles in Figure 2(a), in which $H$ is the mean curvature of the conical surface where the oxygen atom sets. Since the potential decreases along with the increase of the curvature, the water molecule is subject to a force toward the direction of increasing the curvature. Three different cones with the same contact angle $\theta = 50.7°$ but different half-apex angles $\alpha = 41.8°$, $30.0°$ and $9.6°$ are also investigated. The calculated potential-curvature relations are plotted by the other types of dots in Figure 2(a), showing perfect coincidence with each other. The above results reveal that the potential of the water molecule is determined by the curvature of the cone, while independent of the cone half-apex angle.

Further, we show that the above dependence of the potential on curvature is generally valid. Suppose that a particle is interacting with atoms of a curved monolayer material such as graphene through a particle-atom pair distance potential (for example, a van der Waals potential). If the distance, $d$, of this particle to the monolayer surface is smaller than the two principal curvature radii $1/|k_1|$ and $1/|k_2|$ of the surface, then the total potential between the particle and the surface can be approximated as (see SI for detail):

$$U = U_0 \left(1 - 2Hd\right)^{-1/2}, \qquad (1)$$

where $U_0$ denotes the total potential of the particle to the surface as it would be flat and $H = (k_1 + k_2)/2$ is the mean curvature. Since $d$ is usually a fraction of one nanometer, therefore for most practical cases where the curvature radii are much larger than $d$ one can have

$$U = U_0 \left(1 + Hd\right), \qquad (2)$$

The result (1) reveals a new field force that is proportional to the curvature gradient. Since $U_0$ is generally negative regardless of contact angle, the curvature gradient force is always pointing the direction of increasing curvature, independent of wettability.

Since any particle-pair interaction potential is basically an assumption, we consider a more credible approach to the total free energies of a water droplet at different locations on a conical surface. For a liquid droplet with any assigned volume and contact angle on the inner or outer conical surface, using an open resource finite element approach, the Surface Evolver[18], we can precisely calculate its shape through minimizing the free energy $U=\gamma(A-A_0\cos\theta)$. Here, $A$ and $\gamma$ are the liquid-vapor interfacial area and surface energy of the drop and $A_0$ is the solid-liquid interfacial area[19]. We plot in Figure 3 by solid or hollow squares the calculated free energies of a 0.24 $\mu$l water droplet (with a diameter of 0.77 mm as the drop would be spherical) at the outer or inner surface of a cone versus the mean curvature $H$. Three different contact angles 120°, 90°, and 75°, are considered, showing the same tendency of decreasing the free energy with the increase of the curvature

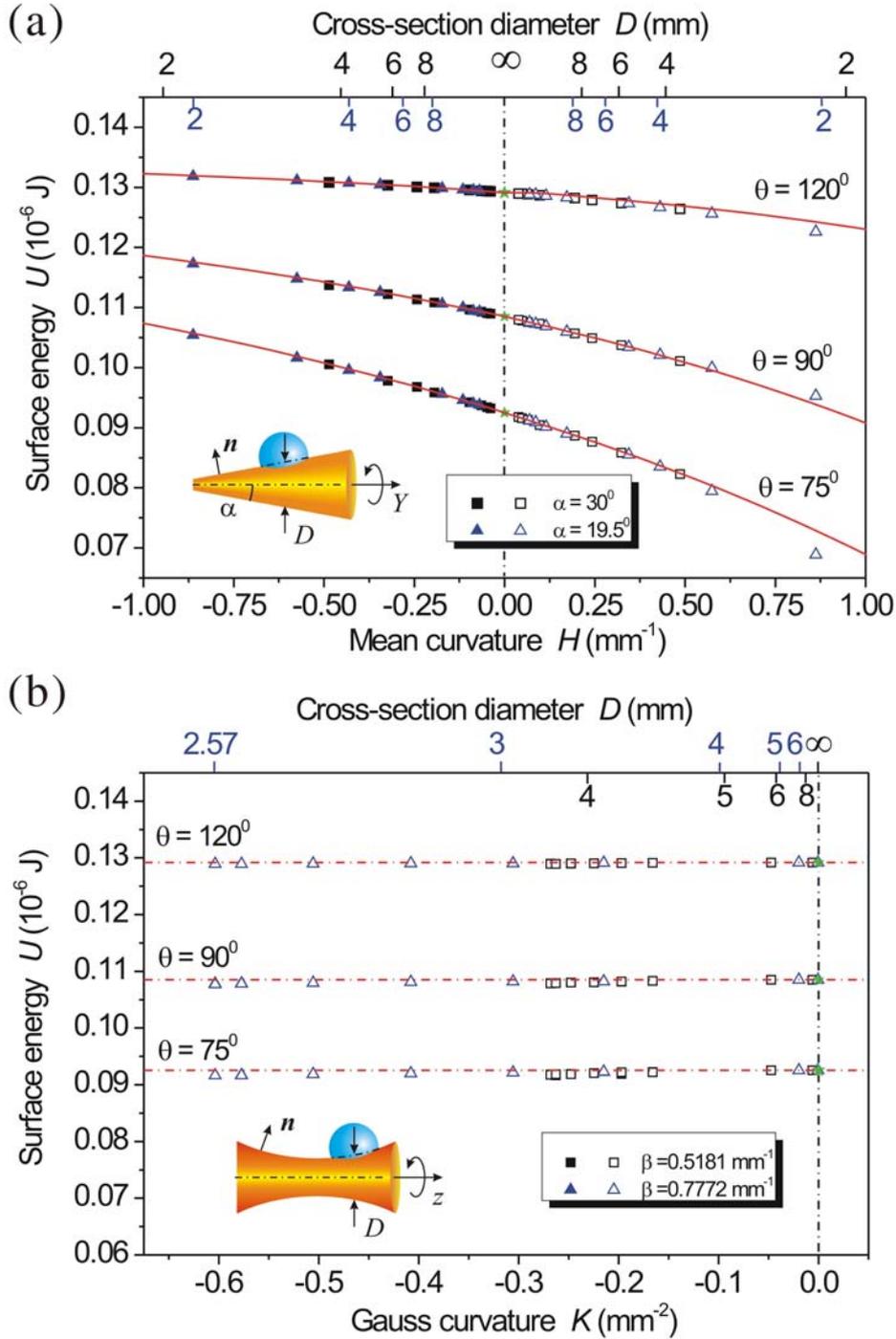

**Figuire 3.** Dependence of the free energies upon curvatures for a water droplet of volume 0.24μL at conical and sperical surfaces. The square and triangle dots plot the numerical results using the Surface Evolver for (**a**) conical surfaces with halp-apex angles $\alpha$ = 30° and 19.5°, respectively, and for (**b**) catenoid surfaces with $\beta$ = 0.5181mm$^{-1}$ and 0.7772 mm$^{-1}$, respectively. The solid red lines in (**a**) represent the theoretical results for the water droplet sitting at spherical surfaces, and the dotted-lines in (**b**) are horizontal that passe through the results for flat surfaces. Each line corresponds to one of the three contact angles $\theta$ = 120°, 90° and 75°. The black and blue labels of cross-section $D$ represent different halp-apex angles $\alpha$ and different $\beta$ in (**a**) and (**b**) respectively.

Since the cone curvature radii corresponding to the results plotted in Figure 3(a) are of millimeters that are six orders in magnitude larger than the van der Waals interaction distance $d$, the results (1) and (2) have the implications that the free energy should substantially be a linear function of the mean cuvrature $H$ and independent of the Gauss' curvature. To valify the latter, we consider a water droplet of volume $0.24 \mu l$ at catenoid surfaces $\cosh(\beta z) = \beta\sqrt{x^2 + y^2}$, that have always zero mean curvature ($H$ = 0) and negative Gauss' curvature $K = -\beta^2 \cosh^{-4}(\beta z)$. The triangle and square dots show the free energies caculated by using the Surface Evolver in which the droplet is on the catenoid surface with $\beta$ = 0.5181mm$^{-1}$ and 0.7772 mm$^{-1}$, respectively, compared with the horizontal dashed-lines that pass through the dots at $K$ = 0. The results shown in Figure 3(b) verify the independence of the free energy from the Gauss' curvature $K$. Furthermore, we find that the departure from linearity of the free energy upon $H$ shown in Figure 3(a) is caused by the variational contact area of the droplet on the conical surface.

The independence of the free energy $U$ upon the Gauss' curvature enables us to give the explicit solution of $U$ as a function of $H$ by analyzing a droplet at spherical surfaces with different radii $1/H$, seeing the Supplementary Information for details. The solid line in Figure 3(a) plots this solution with respect to the same volume (0.24 $\mu$L) water droplet and the three same contact angles as those of the dots, showing perfect agreement with the dots. This explicit solution can be a basis for analytically discuss self-motion behavior of droplets on surfaces with very small curvature radii.

The results shown in Figure 1 indicate that the nanodrop at the nanoscale cone surfaces can reach self-motion speeds of 67 or 110 m/s at outer or inner surfaces of a nanoscale cone, much faster than records[4,20]. It is known however that contact angle hysteresis can result in friction against droplet motion, and thus decrease the self-motion speed or even pin droplets. Since hydrophilic surfaces have usually much larger hysteresis than hydrophobic ones, it is possible that self-motion of droplets may easier take place at a hydrophobic surface than hydrophilic.

Indeed, our preliminary experimental results using glass conical tubes confirm the existence of curvature gradient force at hydrophobic surfaces. Figure 4 shows some selected frames of a water droplet of volume 0.24 $\mu$L at a glass conical tube with half-apex angle of $\alpha$ = 5° coated with a layer of FOTS (1H,1H,2H,2H-perfluorooctyltrichlorosilane)[21]. The humidity and the temperature were 70% and 30°C, respectively. As shown in Figure 4, after being extruded out of the openned small end of the tube ($t$ = 0.2s), the droplet was quickly moving along the tube to a distance ($t$ = 2s) before it stopped and stayed there stilly ($t$ = 6~21s). The contact angle of the still drop at the larger tube section is about $\theta_a$ = 109° while that at the smaller tube section is about $\theta_r$ = 94°. These data show that the coated surface is hydrophobic with the contact angle hysteresis $\theta_a - \theta_r$ = 15°. Thus, the water surface tension along the solid-water-air three-phase contact line will result in a drag force upon the droplet in the direction toward the small tube section. The appearance of this drag force uncovers the existence of the same magnitude driving force toward the larger cone section.

Higher temperature or vibration were often used to overcome the moving resistance due to contact angle hysteresis[4,20]. After applying small lateral vibration, we observed droplet motion toward the larger cone section (Figure 4, $t$ = 21s ~ 34s, or Supplementary Information). The average self-moving speed from time $t$ = 21s to 34s is estimated as 0.12 mm/s.

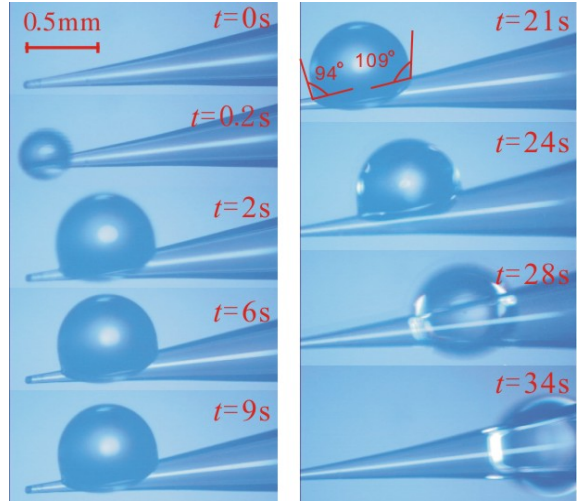

**Figure 4.** Self-motion behavior of a water droplet of volume 0.24 μL at a glass conical tube coated with a layer of FOTS. After being extruded out of the openned small end of the tube ($t$ = 0.2s), the droplet was quickly moving along the tube to a distance ($t$ = 2s) before it stopped and stayed there stilly ($t$ = 6~21s). The advancing and receding contact angles are measured to be about $\theta_a$ = 109° and $\theta_r$ = 94°. The droplet was moving ($t$ = 21s ~ 34s) when the tube was under a small lateral vibration.

It was shown that contact angle gradient can result in small water droplets (0.1 to 0.3 millimeter) in fast self-motion[4,22]. If the contact angle hysteresis is negligible, then the ultimate speed attainable by a small water droplet at a flat surface with contact angles varying from 0° to 180° is $V_{cag} = \sqrt{6\gamma/(\rho r)}$, where $\gamma$, $\rho$, and $r$ are the water-air interfacial energy, water mass density, and droplet radius. In comparison, the ultimate speed attainable by a small water droplet at the outer surface of a cone is $V_{scg} = \sqrt{6\eta\gamma/(\rho R)}$, where $R$ is the cuvature radius of the cone at where the droplet was starting to self-move, and

$$\eta = \frac{(1+\cos\theta)(1+3\cos^2\theta)}{2(1-\cos\theta)(2+\cos\theta)}, \tag{3}$$

which monotonusly decreases from ∞ to 1.00, and then to 0 as θ increases from $0^0$ to $60.8^0$, and then to $180^0$. Since $V_{cag}$ depends upon the droplet size $r$ while $V_{scg}$ is independent of $r$ but depends upon $R$, the latter can be orders higher than the former as $R$ can be easily designed in nanoscale while $r$ is usualy in micro or even millimeters for pratical applications. The above results show that nanoscale curvature gradient surfaces can be fast paths for droplet transfering.

## Acknowledgements

Financial support from the NSFC under grant No.10872114, No.10672089 and No. 10832005 is gratefully acknowledged.

## References

1. Ichimura, K., Oh, S.K. & Nakagawa, M. Light-driven motion of liquids on a photoresponsive surface. *Science* **288**, 1624-1626 (2000).

2. Sumina, Y., Magome, N., Hamada, T. & Yoshikawa, K. Self-running droplet: Emergence of regular motion from nonequilibrium noise. *Phys Rev Lett*. **94**, 068301 (2005).

3. Linke, H. *et al*. Self-propelled Leidenfrost droplets. *Phys Rev Lett*. **96**, 154502 (2006).

4. Khoo, H.S. & Tseng, F.G. Spontaneous high-speed transport of subnanoliter water droplet on gradient nanotextured surfaces. *Applied Physics Letters* **95**, 063108 (2009).

5. Zheng, Y. *et al*. Directional water collection on wetted spider silk. *Nature* **463**, 640-643 (2010).

6. Lorenceau, É. & Quéré, D. Drops on a conical wire. *J. Fluid Mech*. **510**, 29-45 (2004).

7. Liu, J., Xia, R., Li, B. & Feng, X.Q. Directional motion of droplets in a conical tube or on a conical fibre. *Chin. Phys. Lett*. **24**, 3210-3213 (2007).

8. Renvoisé, P., Bush, J.W.M., Prakash, M. & Quéré, D. Drop propulsion in tapered tubes. *Europhys Lett*. **86**, 64003 (2009).

9. de Gennes, P.-G., Brochard-Wyart, F. & Quéré, D. Capillarity and wetting phenomena. Spring, New York (2003).

10. McHale, G., Newton, M.I. & Carroll, B.J. The shape and stability of small liquid drops on fibers. *Oil & Gas Science and Technology* **56**, 47-54 (2001).

11. Plimpton, S. Fast parallel algorithms for short-range molecular-dynamics. *Journal of Computational Physics* **117**, 1-19 (1995).

12. Berendsen, H.J.C., Grigera, J.R. & Straatsma, T.P. The missing term in effective pair potentials. *J. Phys. Chem*. **91**, 6269-6271 (1987).

13. Vega, C. & de Miguel, E. Surface tension of the most popular models of water by using the test-area simulation method. *Journal of Chemical Physics* **126**, 154707 (2007).

14. Allen, D. & Tildesley, D.J. Computer Simulation of Liquid. *Clarendon Press*, Oxford, **1987**.

15. Werder, T., Halther, J.H., Jaffe, R.L., Halicioglu, T. & Koumoutsakos, P. On the water-carbon interaction for use in molecular dynamics simulations of graphite and carbon nanotubes. *Journal of Physical Chemistry B* **107**, 1345-1352 (2003).

16. Nose, S. A molecular-dynamics method for simulations in the canonical ensemble. *Mol. Phys*. **52**, 255-268 (1984).

17. Hoover, W.G. Canonical dynamics-equilibrium phase-space distributions. *Phys Rev A*. **31**, 1695-1697 (1985).

18. Brakke, K.A. Surface Evolver. *www.susqu.edu/brakke/evolver/evolver.html* (2000).

19. Young, T. R. An essay on the cohesion of fluids. *Trans. R. Soc. London*. **95**, 65-87 (1805).

20. Daniel, S., Chaudhury, M.K. & Chen, J.C. Fast drop movements resulting from the phase change on a gradient surface. *Science* **291**, 633-636 (2001).

21. Hsieh, H.Y. et al. Effictive enhancemant of fluorescence detection efficiency in protein microarray assays: Application of a highly fluorinated organosilane as the blocking agent on the background surface by a facile vapor-phase deposition process. *Anal. Chem*. **81**, 7908-916 (2009).

22. Sandhu, A. Nanofluidics: Wedge driven. *Nature Nanotechnology*, Published online: 28 August 2009 | doi:10.1038/nnano.2009.275 (2009).

# Supplementary information

For "Curvature Gradient Driving Droplets in Fast Motion"
by C.J.Lv, C. Chen, Y.J. Yin, F.G. Tseng, Q.S. Zheng

**S.1 Total van der Waals potential energy between a single atom and a general curved surface**

Consider a surface $S$ that is consisting of mono-layered uniform atoms and a single atom M at distance $d$ to the surface $S$. As illustrated in Figure S1, we denote by O the projection point of M to $S$, by $S_O$ the tangent plane of $S$ at O, and by $\{x,y,z\}$ a Cartesian coordinate system such that the origin is at O, the $x$- and $y$-axes are tangent to the two principle curvature lines of $S$, and the $z$-axis is pointing the atom M.

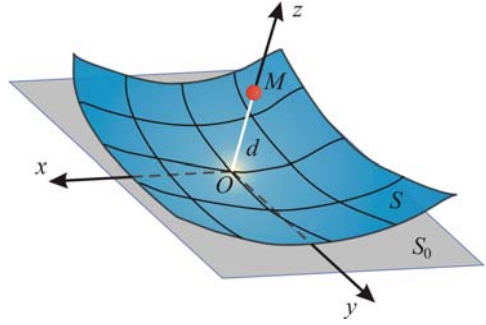

**Figure S1.** The local cartansian coordinate system $\{x,y,z\}$ to depict the interaction between a particle and a general curved surface, with $x$-axis and $y$-axis tangential respectively to the principle curvature lines at the point $O$, $z$-axis normal to the surface, and $d$ denoting the $z$-coordinate of the particle.

By definition, if we use $z = f(x,y)$ to represent the surface $S$, then we have $f(0,0) = f_{,x}(0,0) = f_{,y}(0,0) = f_{,xy}(0,0) = 0$ and

$$f(x,y) \approx \frac{1}{2}(k_1 x^2 + k_2 y^2) \qquad (S1)$$

at all points near O, where $f_{,x}$ and $f_{,y}$ denote the partial derivatives of $f(x,y)$ to $x$ and $y$, respectively, and

$$k_1 = f_{,xx}(0,0), \quad k_2 = f_{,yy}(0,0) \qquad (S2)$$

are the two principle curvatures of the surface at O. The two principle curvature radii are denoted by $R_1 = 1/|k_1|$ and $R_2 = 1/|k_2|$.

Introduce $R = \min\{R_1, R_2\}$ and $S_\delta$ a circular domain of the plane $S_O$ centered at O with radius $\delta$. If $\delta$ is of the same order of $d$ and if $d/R \ll 1$, then the distance $r$ between M and any point of $S$ within $S_\delta$ can be approximated in terms of the following relation:

$$\begin{aligned} r^2 &= x^2 + y^2 + [d - f(x,y)]^2 \\ &\approx x^2 + y^2 + d^2 - 2f(x,y)d \\ &\approx X^2 + Y^2 + d^2 \end{aligned} \tag{S3}$$

by ignoring $O((d/R)^2)$, where

$$X = (1 - k_1 d)^{1/2} x, \quad Y = (1 - k_2 d)^{1/2} y. \tag{S4}$$

Further, the area element at any point within $S_\delta$ can be approximated as

$$\begin{aligned} dS &= (1 + f_{,x}^2 + f_{,y}^2)^{1/2} dxdy \\ &\approx dxdy \\ &\approx [(1 - k_1 d)(1 - k_2 d)]^{-1/2} dXdY \\ &= (1 - 2Hd + Kd^2)^{-1/2} dXdY \end{aligned} \tag{S5}$$

by ignoring a relative error of the order $O((\delta/R)^2)$, where $H = (k_1 + k_2)/2$ and $K = k_1 k_2$ are the mean and Gauss' curvature.

Denote by $\phi(r_i)$ the van der Waals interaction potential between M and the $i$th atom on $S$ which is at distance $r_i$ to M. Then, the total potential is

$$U(d) = \sum_i \phi(r_i) \approx \rho \int_S \phi(r) dS, \tag{S6}$$

where $\rho$ is the density of surface atoms, i.e. the number of surface atoms per unite area, and is assumed to be constant. By noting that van der Waals interactions are typically negligible as the distance is larger than 1nm, we further have the following approximation:

$$U(d) \approx \rho \int_{S_\delta} \phi(r) dxdy, \tag{S7}$$

if $d$ is of the order of 1nm. Substituting (S5) into (S7) yields

$$\begin{aligned} U(d) &\approx \rho \int_{S_\delta} \phi(\sqrt{X^2 + Y^2 + d^2}) dxdy \\ &\approx U_0(d)(1 - 2Hd + Kd^2)^{-1/2} \end{aligned} \tag{S8}$$

where

$$U_0(d) = \rho \int_{S_O} \phi(\sqrt{X^2 + Y^2 + d^2}) dXdY \tag{S9}$$

is the total van der Waals energy of M as if S would be flat. Since $Kd^2$ is also of the order $O((\delta/R)^2)$, if we ignore this term, we finally obtain Eq.(1), namely.

$$U(d) = U_0(d)(1-2Hd)^{-1/2} \tag{S10}$$

## S.2 Ultimate speed attainable by a small water droplet on a curved and a flat surface

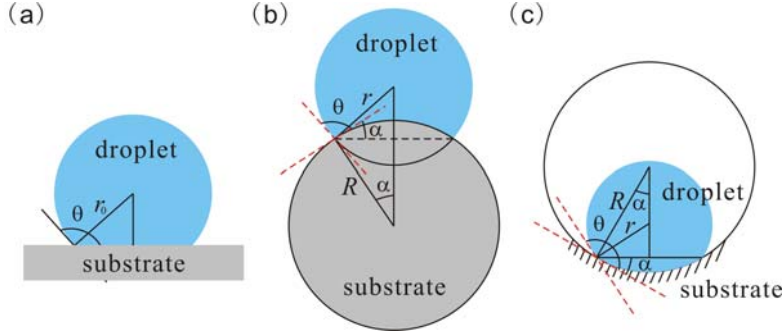

**Figure S2.** Sketch about the wetting states of the same water droplet on a flat **(a)**, outer **(b)** and inner **(c)** of a curved surfaces, respectively.

For a water droplet with contact angle $\theta$ and volume $V_0$, it's not difficult to obtain the follow relationships (Figure S2**(a)**):

$$V_0 = \frac{\pi}{3} r_0^3 \left(2 - 3\cos\theta + \cos^3\theta\right), \tag{S11}$$

$$E_0 r_0 = 3V_0 \gamma, \tag{S12}$$

where $r_0$ is the radii of the spherical of the water droplet on a flat surface, $E_0$ is the surface tension energy of the water droplet, such as $E_0 = \gamma(A_{LV} - A_{SL} \cdot \cos\theta)$. $A_{LV}$ and $\gamma$ are the liquid-vapor interfacial area and surface energy of the drop and $A_{SL}$ is the solid-liquid interfacial area. When the water droplet is sitting on the outer of a spherical surface (Figure 2(b)), we can get the following relationships:

$$\begin{cases} V_0 = V_1 - V_2 \\ R\sin\alpha = r\sin(\theta + \alpha) \end{cases}, \tag{S13}$$

here, $r$ is the radius of the water droplet on the spherical surface, $R$ is the radius of the contact area, $\alpha$ is shown in Figure S2**(b)** and

$$V_1 = \frac{\pi}{3} r^3 \left[1 - \cos(\theta + \alpha)\right]^2 \left[2 + \cos(\theta + \alpha)\right], \tag{S14}$$

$$V_2 = \frac{\pi}{3} R^3 (1 - \cos\alpha)^2 (2 + \cos\alpha). \tag{S15}$$

When the radii of the curved spherical surface is larger enough, for example, $r \ll R$, $\alpha \ll 1$, we introduce $\varepsilon_0 = r_0/R$, and let

$$r = r_0 (1 + k\varepsilon_0). \tag{S16}$$

$k$ is an undetermined parameter in Eq (S16). By Taylor series, we can expand $\cos\alpha$ and $\sin\alpha$ to $\cos\alpha=1-\alpha^2/2+\alpha^4/24+\cdots$ and $\sin\alpha=\alpha-\alpha^3/6+\cdots$, and then we can get:

$$V_1 \approx \frac{\pi}{3}r^3\left[\left(2-3\cos\theta+\cos^3\theta\right)+3\alpha\sin^3\theta\right], \tag{S17}$$

$$V_2 \approx \frac{\pi}{4}\alpha^4 R^3, \tag{S18}$$

$$A_{LV} = 2\pi r^2\left[1-\cos(\theta+\alpha)\right] \approx 2\pi r^2\left(1-\cos\theta+\alpha\sin\theta\right), \tag{S19}$$

$$A_{SL} = 2\pi R^2\left(1-\cos\alpha\right) \approx \pi\alpha^2 R^2, \tag{S20}$$

By Eq (S13) and ignoring a relative error of the order $O(\varepsilon_0^2)$, we can further get:

$$k = -\frac{3}{4} \cdot \frac{\sin^4\theta}{2-3\cos\theta+\cos^3\theta}, \tag{S21}$$

and the surface energy of a water droplet on the outer of the spherical surface (Figure S2(**b**)):

$$E = E_0\left[1+2kr_0\left(1-\frac{4}{3}\cdot\frac{1}{\sin^2\theta}\right)H\right], \tag{S22}$$

here, $H=1/|R|$ is the mean curvature of the spherical surface. When the water droplet is sitting on the inner of the spherical surface (Figure S2(**c**)), we can also get the same expression of Eq (S22), but $H=-1/|R|$ in this case. In other words, for arbitrary curved surfaces, whether $H$ is positive or negative is determined by the shape of the contact area.

So far, we can see that the energy of a water droplet on a curved surface is also determined by the local curvature of the contact area of the curved surface, and this relationship is never appeared in the past. From Figure 3, this conclusion is further validated. We can see the energy will be equivalent with each other no matter the curved surface is a spherical surface or a conical surface (or an arbitrary curved surface) if the local curvature of the contact area has a certain value.

We will give the ultimate speed which is driven by the curvature gradient of the curved surface (the gradient of the mean curvature $H$) along the direction of the movement (for example, in Figure 1(**a**), $s$ is the movement direction of the water droplet, but below, $H$ and $s$ represent the mean curvature and movement direction for a water droplet on an arbitrary curved surface). As we know, the net force can be wrote as $F=-\partial E/\partial s$, by Eq (S22), we can get:

$$-2kE_0 r_0\left(1-\frac{4}{3}\cdot\frac{1}{\sin^2\theta}\right)\cdot\frac{\partial H}{\partial s} = \rho V_0 \ddot{s}, \tag{S23}$$

here, the dot "·" means the time derivative of $s$, by some calculations and integrations,

it is not difficult to get:

$$\dot{s}^2 = \frac{6\eta\gamma}{\rho}\left(\frac{1}{R_0} - \frac{1}{R}\right), \tag{S24}$$

here, $R_0$ and $R$ are the radii of the contact area of the curved surface on which the droplet starts to move and it will pass by. If $R$ is large enough, for example $R\to\infty$, we can get the ultimate speed attainable of the water droplet on the curved surface $V_{scg} = \dot{s}|_{R\to\infty} = \sqrt{6\eta\gamma/(\rho R_0)}$, here:

$$\eta = \frac{(1+\cos\theta)(1+3\cos^2\theta)}{2(1-\cos\theta)(2+\cos\theta)}. \tag{3}$$

If a water droplet is sitting on a flat gradient surface, it will move from the direction with large contact angle to the direction with small contact angle, we can get the net force $F = -\partial E/\partial s$ along the moving direction $x$. By the relationships $E_0 = 3V_0\gamma/r_0$ and $E = 3V_0\gamma/r$, we can get:

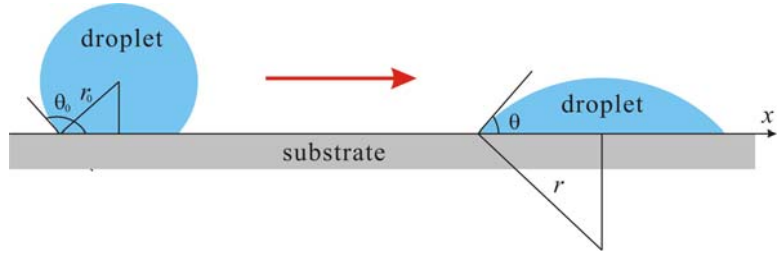

**Figure S3.** Wetting behaviors of a water droplet on a flat gradient surface.

$$-3V_0\gamma\frac{\partial}{\partial x}\left(\frac{1}{r}\right) = \rho V_0 \ddot{x}. \tag{S25}$$

After some calculations and integrations, we obtain:

$$(\dot{x})^2 = \frac{6\gamma}{\rho}\left(\frac{1}{r_0} - \frac{1}{r}\right). \tag{S26}$$

here, $r_0$ and $r$ are the radii of the droplet when it starts to move and during the movement. If the other side of the surface is hydrophilic enough, in other words, if $r\to\infty$, we can get the ultimate speed attainable of the water droplet on the flat surface $V_{cag} = \dot{x}|_{r\to\infty} = \sqrt{6\gamma/(\rho r_0)}$.

**Supplementary Movies:**

**Movie 1** (S1_MD.mov, 8.21 MB) – Molecular dynamics simulations results of water droplet self-motion

The water droplet on the outer/inner conical surface is spontaneously moving toward the larger/smaller cross-section area no matter the value of the water contact angle ($\theta$=50.7° (hydrophilic), 138° (hydrophobic), or nearly 180° (superhydrophobic)). The droplet was bouncing when it meets the boundary.

**Movie 2** (S2_Experiment.mov, 841 KB) - Experiment results with a 0.24 $\mu$l water droplet

After being extruded out of the opened small end of the tube, the droplet was quickly moving along the tube to a distance before it stopped. When we gave a random lateral disturbance, the droplet was moving to the direction with larger cross-section.